\documentclass[prd,twocolumn,showpacs]{revtex4}
\usepackage{amsfonts,amsmath,amssymb,graphicx}

\newcommand{ \be}{\begin{equation}}
\newcommand{ \ee}{\end{equation}}
\newcommand{ \bea}{\begin{eqnarray}}
\newcommand{ \eea}{\end{eqnarray}}
\newcommand{ \eq}[1]{Eq.~(\ref{eq:#1})}
\newcommand{ \bm}   {\boldmath}

\newcommand{ \mysmall}[1]{\scriptscriptstyle #1} 
\newcommand{ \pithr}{\Pi_{\rm\mysmall{thr}}}
\let\oldhat\hat
\renewcommand{\vec}[1]{\mathbf{#1}}
\renewcommand{\hat}[1]{\oldhat{\mathbf{#1}}}

\begin{document}
\noindent
LPN14-049

\title{\bm Positronium contribution to the electron $g$-2}

\author{M.~Fael}
\email{matteo.fael@pd.infn.it}
\affiliation{Dipartimento di Fisica e Astronomia, Universit\`a di Padova}
\affiliation{Istituto Nazionale Fisica Nucleare, Sezione di Padova, I-35131 Padova, Italy}
\affiliation{Physik-Institut, Universit\"{a}t Z\"{u}rich, CH-8057, Z\"{u}rich, Switzerland}

\author{M.~Passera} 
\email{passera@pd.infn.it}
\affiliation{Istituto Nazionale Fisica Nucleare, Sezione di Padova, I-35131 Padova, Italy}

\begin{abstract} 
\noindent 
The contribution of positronium to the electron $g$$-$$2$ ($a_e$) has been computed in Ref.~\cite{Mishima:2013ama} and found to be of the same order of $\alpha$ as that of five-loop perturbative QED. We confirm this result and correct a few errors in its first derivation. As recently calculated in Ref.~\cite{Melnikov:2014lwa}, a continuum nonperturbative contribution to $a_e$ cancels one-half of the positronium one. We show by explicit calculation that the remaining half is already included in the five-loop perturbative result. We also show that it arises from the class I(i) of five-loop diagrams containing only one closed electron loop.
\end{abstract} 

\date{\today}
\maketitle

\section{Introduction}
\label{sec:Introduction}
 
The leading contribution of positronium, the $e^+$$e^-$ bound state, to the anomalous magnetic moment of the electron $(a_e)$ has been computed in Ref.~\cite{Mishima:2013ama}. The result of this calculation,
\be
	a_e^{\mysmall \rm  P} = \frac{\alpha^5}{4\pi} \zeta(3) \left( 8\ln2 - \frac{11}{2} \right) =  0.89 \times 10^{-13},
\label{eq:aeP}
\ee
where $\zeta(3) = 1.202\ldots$ and $\alpha$ is the fine-structure constant, is of the same order of $\alpha$ as the perturbative QED five-loop contribution $a_e^{\mysmall (10)} = 9.16 \, (58) \left( \alpha/\pi \right)^5$~\cite{Aoyama:2012wj}. This bound-state contribution is also comparable with the electroweak one, $a_{e}^{\mysmall \rm EW} = 0.2973 \, (52) \times 10^{-13}$~\cite{Mohr:2012tt,EW}, and with the present experimental uncertainty of $a_e$, $2.8 \times 10^{-13}$~\cite{Gabrielse}. It seems reasonable to expect a reduction of this experimental error to a part in $10^{-13}$ (or better) in ongoing efforts to improve the measurement of the electron (and positron) anomalous magnetic moment~\cite{GabrielseLDMBook}. Work is also in progress to reduce the error induced by the uncertainty of $\alpha$ in the theoretical prediction for $a_e$~\cite{Bouchendira:2010es}.

A test of the electron $g$$-$$2$  at the level of $10^{-13}$ (or below) is therefore a goal that may be achieved not too far in the future with ongoing experimental work. This will bring $a_e$ to play a pivotal role in probing new physics~\cite{Giudice:2012ms}. It will also provide the opportunity to test whether the long-standing 3--4$\sigma$ discrepancy $\Delta a_{\mu}$ in the muon $g$$-$$2$ manifests itself in the electron one~\cite{Giudice:2012ms,Terranova:2013vfa}. In fact, as shown in Ref.~\cite{Giudice:2012ms}, in a large class of new-physics models, new contributions to lepton magnetic moments scale with the square of the lepton masses, so that the anomaly in $\Delta a_{\mu}$ suggests a new-physics effect in $a_e$ of $(0.7 \pm 0.2) \times 10^{-13}$, a value comparable with $a_e^{\mysmall \rm  P}$. A check of~\eq{aeP} is therefore clearly warranted. This is presented in Sec.~\ref{sec:Poles}, where we confirm the result of~\eq{aeP} and correct a few errors in its derivation in Ref.~\cite{Mishima:2013ama}.

Recently, the authors of Ref.~\cite{Melnikov:2014lwa} pointed out the presence of the continuum nonperturbative contribution 
\be
	a_e({\rm vp})^{\rm cont, np} = - \frac{|\alpha|^5}{8\pi} \zeta(3) \left( 8\ln2 - \frac{11}{2} \right)
\label{eq:aeC}
\ee
arising from the region right above the $s=4m^2$ threshold, which corresponds to $e^+ e^-$ scattering states with the exchange of Coulomb photons. Comparing Eqs.~(\ref{eq:aeP}) and (\ref{eq:aeC}) they showed that this additional ${\cal O}(\alpha^5)$ nonperturbative contribution cancels one-half of that of the positronium poles. The question is therefore how to deal with the remaining half: should one add it to the perturbative five-loop QED result of Ref.~\cite{Aoyama:2012wj}? Reference~\cite{Melnikov:2014lwa} argued that this remaining $a_e^{\mysmall \rm  P}/2$ term is already contained in the perturbative ${\cal O}(\alpha^5)$ contribution to $a_e$ computed in Ref.~\cite{Aoyama:2012wj} and, therefore, it should not be added to it. On the other hand, one of the authors of the five-loop calculation in~\cite{Aoyama:2012wj} has recently claimed that positronium contributes to $a_e$ only through diagrams of ${\cal O}(\alpha^7)$ or higher~\cite{Hayakawa:2014tla}. Also, on more general grounds~\cite{Braun1968}, Ref.~\cite{Eides:2014swa} argued that $a_e^{\mysmall \rm  P}$ simply does not exist.

In order to clarify this point, in Sec.~\ref{sec:Threshold} we first use the closed form for the QED vacuum polarization function near the $s=4m^2$ threshold of Refs.~\cite{Braun1968,BCR1973} to verify that the total (positronium poles plus continuum) nonperturbative contribution to $a_e$ arising from the threshold region is equal to $a_e^{\mysmall \rm  P}/2$. Then, using the analytic QED vacuum polarization at four-loop recently computed in Ref.~\cite{Baikov:2013ula}, we show explicitly that the perturbative five-loop calculation of $a_e$ of Ref.~\cite{Aoyama:2012wj} does indeed contain the remaining term $a_e^{\mysmall \rm  P}/2$, in agreement with the arguments of Ref.~\cite{Melnikov:2014lwa}. 
Conclusions are drawn in Sec.~\ref{sec:Conclusions}.

\section{Positronium Poles}
\label{sec:Poles}

Let us consider QED with only electrons and photons. The vacuum polarization tensor is given by
\begin{align}
	i \Pi^{\mu \nu}(q) & = i \Pi(q^2) \left( g^{\mu \nu}q^2 -q^{\mu}q^{\nu} \right) \nonumber \\
	& = \int d^4 x \, {\rm e}^{iqx} \langle 0| T\{j^{\mu}(x) \, j^{\nu}(0)\} |0 \rangle,
\label{eq:Pimunu}
\end{align}
where $j^{\mu}(x) = -e \bar{\psi}(x) \gamma^{\mu} \psi(x)$ is the electromagnetic current. In perturbative calculations, $\Pi(q^2)$ is analytic in the complex $q^2$-plane except for cuts along the positive real axis beginning at $q^2=(2lm)^2$, where $m$ is the electron mass and $l=0,1,2,\ldots$. The $q^2=0$ branch-point $(l=0)$ is the threshold value for production of three (or a higher odd number of) real photons, while $l=1$ corresponds to the threshold for the creation of a real $e^+ e^-$ pair by a virtual photon.

An electron-positron bound state will appear as an additional pole singularity in $\Pi(q^2)$ below the $q^2=(2m)^2$ branch-point. In fact, there is an infinite number of such poles, each corresponding to an energy state of positronium. In any of its $n$  discrete states ($n=1,2,3,\ldots$ is the principal quantum number), positronium may be regarded as an (unstable) particle with mass $M_n = 2m-{\cal E}_n$, where ${\cal E}_n>0$ is the binding energy. To leading order in $\alpha$, ${\cal E}_n = m \alpha^2/4n^2$~\cite{Landau4}. To compute the leading-order contribution of positronium to $a_e$ we can use the approximation $M_n \approx 2m$. Positronium will be treated as a two-particle nonrelativistic bound state.

To determine the contribution of positronium to $\Pi^{\mu \nu}(q)$ in the neighborhood of its poles, we write explicitly the time-ordered product appearing in~\eq{Pimunu}
\begin{align}
	\langle 0| T\{j^{\mu}(x) \, j^{\nu}(0)\} |0 \rangle 	& = 
	\theta(x^0) 	\langle 0| j^{\mu}(x) \, j^{\nu}(0)|0 \rangle \nonumber \\
	& + \theta(-x^0)	\langle 0| j^{\nu}(0) \, j^{\mu}(x)|0 \rangle
\label{eq:Tproduct}
\end{align}
and compute $\langle 0| j^{\mu}(x) \, j^{\nu}(0)|0 \rangle$ by inserting between the two currents the completeness relation 
\be
	(1)_{\mysmall \rm  P} = 
	\sum_{n,\sigma} \int \frac{d^3p}{(2\pi)^3} \frac{1}{2E_{n,\vec{p}}} |n,\vec{p},\sigma \rangle \langle n,\vec{p},\sigma |
\label{eq:completeness}
\ee
for the positronium one-particle states
\be
	|n,\vec{p},\sigma \rangle = 
	\int \frac{d^3k}{(2\pi)^3} \sqrt{\frac{2E_{n,\vec{p}}}{2E_+ 2E_-}} \, \tilde{\phi}_{n,\vec{p}} (\vec{k}) |\vec{k_+}, \vec{k_-}, \sigma \rangle.
\label{eq:positronium}
\ee
In~\eq{completeness}, $\vec{p}$ and $E_{n,\vec{p}} = \sqrt{\vec{p}^2 +M_n^2}$ are the three-momentum and energy of positronium,  and $\sigma$ indicates its four spin states: three spin-1 states (triplet) and one spin-0 state (singlet). In~\eq{positronium}, positronium states have been expressed as a linear superposition of free $e^+$ and $e^-$ states with three-momenta $\vec{k}_{\pm}$, respectively, and energies $E_{\mysmall \pm} = \sqrt{ \vec{k}_{\mysmall \pm}^2 +m^2  }$, with $\vec{p} = \vec{k}_+ + \vec{k}_-$ and $\vec{k} = (\vec{k}_+ - \vec{k}_-)/2$. This superposition is weighted by the momentum-space Coulomb wavefunction $\tilde{\phi}_{n,\vec{p}} (\vec{k})$, which gives the amplitude for finding a particular value of $\vec{k}$ for a positronium state $n$ with total momentum $\vec{p}$. In the nonrelativistic bound-state approximation employed in this paper, 
$|\vec{k}| \sim {\cal O} (\alpha m ) \ll  m$~\cite{PS,IZ}.

Our result for the positronium contribution to $\langle 0| j^{\mu}(x) \, j^{\nu}(0)|0 \rangle$ is
\begin{align}
& \langle 0| j^{\mu}(x) \, j^{\nu}(0)|0 \rangle_{\mysmall \rm  P} = -16 \pi \alpha \,  \times   \nonumber \\
& \times \sum_n  \frac{|\phi_{n,0}(0)|^2}{M_n}  
\int \frac{d^3p}{(2\pi)^3} \frac{e^{-ipx}}{2E_{n,\vec{p}}} \left( M_n^2 g^{\mu \nu} -p^{\mu} p^{\nu} \right),
\label{eq:jmujnu}
\end{align}
where $p^{\mu}=(E_{n,\vec{p}},\vec{p})$ and $\phi_{n,0}(0)$ is the position-space wavefunction at the origin in the rest frame of positronium. Our result in~\eq{jmujnu} differs from that in Eq.~(6) of Ref.~\cite{Mishima:2013ama}. Ours has an additional factor 
\be
	\xi_{n,\vec{p}} = - \frac{M_n}{E_{n,\vec{p}}}.
\label{eq:xi}
\ee
Apart from the sign difference, this factor $\xi_{n,\vec{p}}$ renders our expression in~\eq{jmujnu} Lorentz invariant (we note that, in the $|\vec{k}| \ll  m$ limit, the ratio $\phi_{n,0}(0)/\sqrt{M_{n}}$ is a Lorentz scalar under boosts with momentum $\vec{p}$). On the contrary, the result for the positronium contribution to $\langle 0| j^{\mu}(x) \, j^{\nu}(0)|0 \rangle$ of Ref.~\cite{Mishima:2013ama} is not Lorentz invariant.

Contrary to Ref.~\cite{Mishima:2013ama}, \eq{jmujnu} has been obtained summing over all spin states of positronium. However, the spin-0 state (singlet) does not contribute because, in the nonrelativistic bound-state approximation employed, the expression for  $\langle 0| j^{\mu}(x) \, j^{\nu}(0)|0 \rangle_{\mysmall \rm  P}$ has no angular dependence. The $e^+ e^-$ bound state is therefore in an $s$-wave with zero orbital angular momentum, and angular momentum conservation requires that the total spin of the bound state is equal to 1 (triplet).

The leading contribution of positronium to $\Pi(q^2)$ can be immediately obtained from~Eqs.~(\ref{eq:Tproduct}) and (\ref{eq:jmujnu}) using the integral representation 
$
	\theta(t) = -i \int (d \omega /2\pi) e^{i \omega t}/(\omega - i \epsilon)
$
for the step function ($\epsilon>0$). The result is 
\be
	\Pi_{\mysmall \rm  P}(q^2) = -16 \pi \alpha \sum_n  \frac{|\phi_{n,0}(0)|^2}{M_n} \frac{1}{q^2 -M_n^2 +i \epsilon}.
\label{eq:pi}
\ee
Once again, our~\eq{pi} differs from Eq.~(8) in Ref.~\cite{Mishima:2013ama} by a factor $\xi_{n,\vec{q}}$, which renders our result for $\Pi_{\mysmall \rm  P}(q^2)$ Lorentz invariant, while that in~\cite{Mishima:2013ama} is not. Also, the nonrelativistic limit $E_{n,\vec{q}} \to M_n$ taken in~\cite{Mishima:2013ama} to compute the contribution of $\Pi_{\mysmall \rm  P}(q^2)$ to $a_e$ (which should not be confused with the nonrelativistic bound-state approximation $|\vec{k}| \ll  m$) is not tenable. The sign of the residues of the poles in our~\eq{pi},
\be
	Z_n = -16 \pi \alpha \frac{|\phi_{n,0}(0)|^2}{M_n} <0,
\label{eq:Z}
\ee
is in agreement with the sign of the spectral density of the K\"{a}ll\'{e}n-Lehmann representation for $\langle 0| T\{j^{\mu}(x) \, j^{\nu}(0)\} |0 \rangle$~\cite{Jegerlehner:2008zza}. The leading contribution of positronium to the imaginary part of $\Pi(q^2)$ is given by
\be
	{\rm Im} \, \Pi_{\mysmall \rm  P}(q^2) = - \pi \sum_n Z_n \, \delta \! \left(q^2 -M_n^2 \right).
\label{eq:ImPi}
\ee
This result differs from that reported in Ref.~\cite{BCR1973}, ours being twice theirs, while it agrees with that of Ref.~\cite{Hoang:1997vs} obtained via the nonrelativistic Coulomb Green's function~\cite{Hostler:1963zz, Schwinger:1964zzb, Braun1968, Caswell:1978mt} (see also~\eq{impithr} below).

The contribution to $a_e$ of the diagram in Fig.~\ref{fig:vp}, containing the vacuum polarization insertion in the internal photon line of the one-loop electron vertex diagram, can be computed using a (subtracted) dispersion relation for the vacuum polarization.  The result can be cast in the form~\cite{BM61,Jegerlehner:2008zza,Jegerlehner:2009ry}
\be
	a_e({\rm vp}) = \frac{\alpha}{\pi^2} \int_{0}^{\infty} \frac{ds}{s} \,\, {\rm Im} \, \Pi(s+i\epsilon) \, K(s),
\label{eq:aevp}
\ee
where
\be
	K(s) = \int_0^1 dx \, \frac{x^2 (1-x)}{x^2 + (1-x) (s/m^2)}
\ee
is a positive function. The $i\epsilon$ prescription indicates that, in correspondence of a cut, the function ${\rm Im} \, \Pi(s)$ must be evaluated right above it, at $s+i\epsilon$. 
\begin{figure}[h]
\begin{center}
\includegraphics[width=6.5cm]{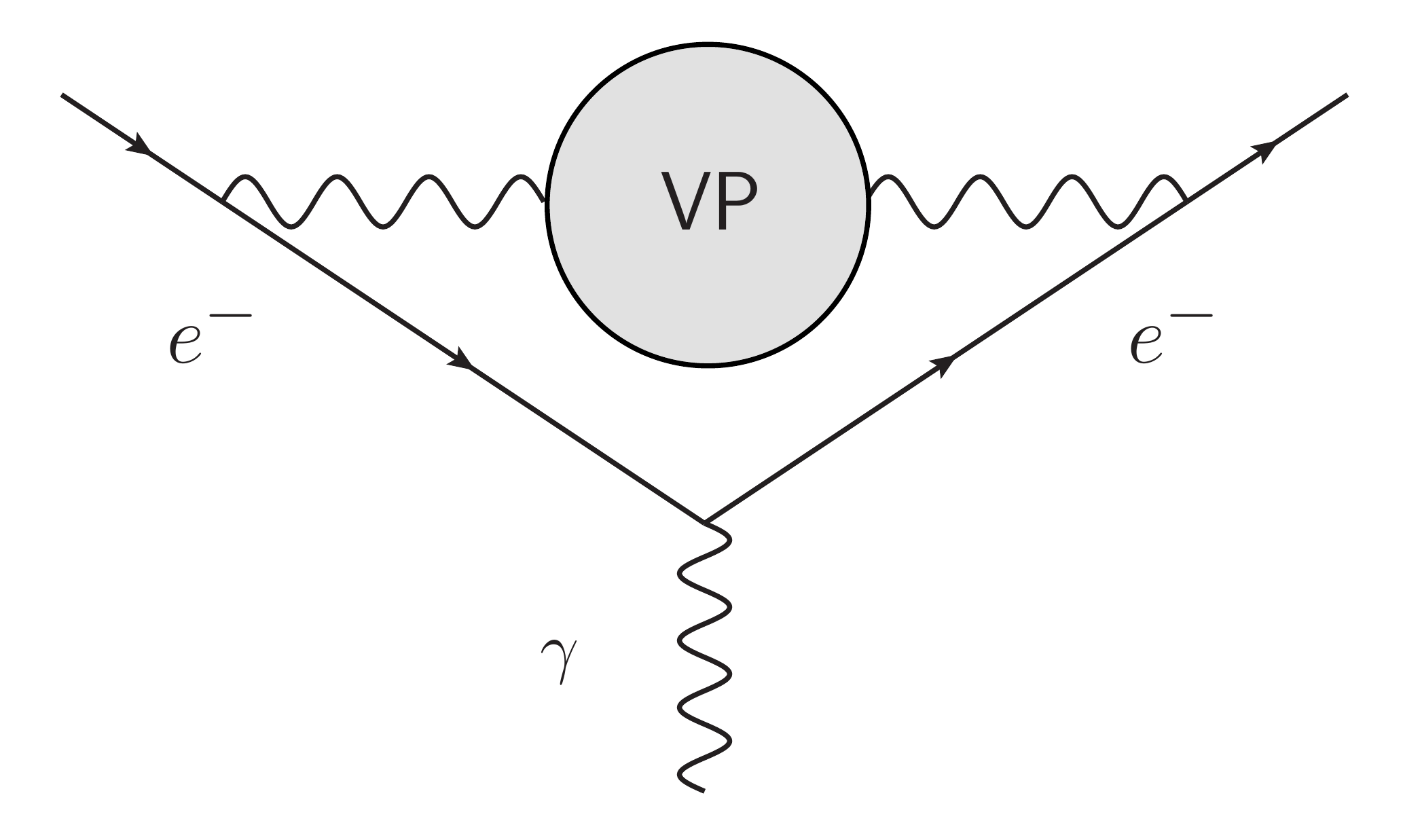}
\caption{Leading vacuum polarization contribution to $a_e$.}
\label{fig:vp}
\end{center}
\end{figure}
Equation~(\ref{eq:aevp}) differs from Eq.~(13) of Ref.~\cite{Mishima:2013ama} by an overall minus sign. This sign can be checked, for example, inserting in~\eq{aevp} the imaginary part of the second-order (one-loop) contribution to $\Pi(q^2)$
\be
	{\rm Im} \, \Pi^{\mysmall (2)}\!(s +i \epsilon) = \theta \! \left(s - 4m^2 \right) \frac{\alpha}{3} 
	\sqrt{ 1 - \frac{4m^2 }{s} }  \! \left( 1+ \frac{2m^2 }{s} \right).
\label{eq:ImPi2}
\ee
One obtains
$
	a_e^{\mysmall (4)} ({\rm vp}) = \left(119/36 - \pi^2/3 \right) \left(\alpha/\pi \right)^2 \!\!,
$
the well-known positive result for the two-loop QED contribution to $a_e$ originated by the one-loop $e^+ e^-$ contribution to the photon self-energy (see e.g.~\cite{Jegerlehner:2009ry,Passera:2004bj}). Similarly, including hadronic effects, the leading-order hadronic contribution to $a_e$ can be obtained via the dispersive integral in~\eq{aevp} with ${\rm Im} \, \Pi_h(s) = s \sigma_h(s)/4\pi \alpha$, where $\sigma_h(s)$ is the total cross section for $e^+e^-$ annihilation into any hadronic state (with vacuum polarization and initial-state QED corrections subtracted off), leading to
$ 
    a_{e}^{\mysmall \rm HLO} = 18.66 \, (11) \times 10^{-13}
$~\cite{Nomura:2012sb,Jegerlehner:2009ry},
once again a positive contribution.

\begin{figure}[h]
\begin{center}
\includegraphics[width=6.5cm]{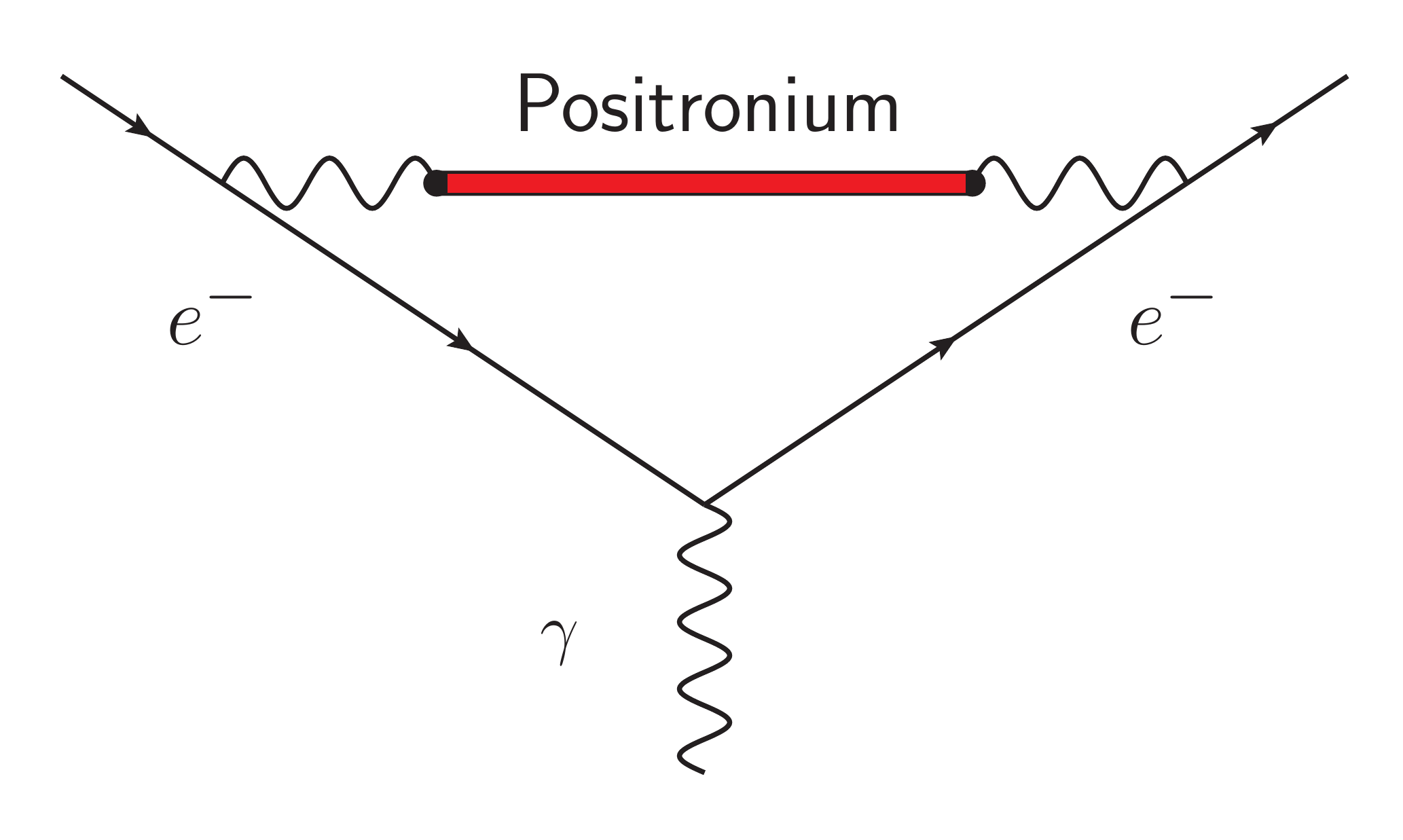}
\caption{Leading contribution of positronium to $a_e$.}
\label{fig:positronium}
\end{center}
\end{figure}

The leading contribution of positronium to $a_e$, depicted in Fig.~\ref{fig:positronium}, can be immediately derived inserting \eq{ImPi} into the integral in~\eq{aevp}. Using the explicit expression for the position-space wavefunction $\phi_{n,0}(0)$ at the origin in the rest frame of positronium~\cite{Landau4}
\be
	|\phi_{n,0}(0)|^2= \frac{(m \alpha)^3}{8 \pi n^3},
\ee
the approximation $M_n \approx 2m$ (thus neglecting terms of {\cal O}$(m \alpha^2)$), and 
$K(4m^2) = 8\ln2 - 11/2$,
we obtain~\eq{aeP}. We note that the Riemann zeta function $\zeta(3) = \sum_{n=1}^\infty 1/n^3$ is due to the sum over the residues of the poles. 
Equation~(\ref{eq:aeP}) can equivalently be computed by direct integration of the Feynman diagram in Fig.~\ref{fig:vp} with the subtracted vacuum polarization function 
\be
	\Pi_{\mysmall \rm  P}(q^2) -\Pi_{\mysmall \rm  P}(0) = 
		\sum_n \,  \frac{Z_n}{M_n^2}  \, \frac{q^2}{q^2 -M_n^2 +i \epsilon}
\ee
without employing its dispersion representation. 

Our result for $a_e^{\mysmall \rm  P}$ agrees with that of Ref.~\cite{Mishima:2013ama}. In fact, the sign error in the calculation of $\langle 0| j^{\mu}(x) \, j^{\nu}(0)|0 \rangle$ in~\cite{Mishima:2013ama} is compensated by the incorrect sign in Eq.~(13) of that reference. Also, as we discussed earlier, the erroneous additional factor $E_{n,\vec{q}}/M_n$ present in Eq.~(8) of Ref.~\cite{Mishima:2013ama} was set to one taking the incorrect limit $E_{n,\vec{q}} \to M_n$. In spite of these shortcomings, Ref.~\cite{Mishima:2013ama} provides the correct contribution of positronium to the $g$$-$$2$ of the electron and was the first, to our knowledge, to compute it.

\section{Threshold Contribution}
\label{sec:Threshold}

In this section we study the nonperturbative contribution to $a_e({\rm vp})$ arising from the region near the electron-positron threshold, both below and above $q^2=4m^2$, and discuss its relation with perturbative QED results.

Let us start considering the vacuum polarization function close to $q^2 \approx 4m^2$ given by~\cite{Braun1968,BCR1973,Hoang:1997vs}
\be
	\pithr(q^2) = \pithr^{\mysmall(2)}(q^2) + \pithr^{\mysmall(4)}(q^2) + A(\beta),
\label{eq:pithr}
\ee
where
\bea	
	\pithr^{\mysmall(2)}(q^2) &=& \alpha \left( \frac{8}{9 \pi} + \frac{i}{2}\beta \right), 
	\label{eq:pithr2} \\
	\pithr^{\mysmall(4)}(q^2) &=& 
	\alpha^2 \left[ \frac{1}{4 \pi^2}\left( 3 - \frac{21}{2} \zeta(3) \right) +\frac{11}{32} \right. \nonumber \\
	&& \left. - \frac{3}{4}\ln 2 -\frac{1}{2}\ln(-i\beta) \right], 
	\label{eq:pithr4} \\
	A(\beta) &=& -\frac{\alpha^2}{2} \left[ \gamma +\psi \!\left( 1- \frac{i \alpha}{2 \beta} \right) \right],
	\label{eq:pithrA}
\eea
$\gamma = 0.577\ldots$ is Euler's constant, $\psi(z) = d \ln \Gamma(z)/dz$ is the digamma function, and
$
	\beta = \sqrt{1- 4m^2/q^2}
$
(for $q^2>4m^2$, $\beta$ corresponds to the velocity of the electron and the positron in their c.m.\ frame). The functions $\pithr^{\mysmall(2)}(q^2)$ and $\pithr^{\mysmall(4)}(q^2)$ are the leading terms of the one- and two-loop functions $\Pi^{\mysmall(2)}(q^2)$ and $\Pi^{\mysmall(4)}(q^2)$, respectively, in the nonrelativistic limit $\beta \to 0$. For example, \eq{ImPi2} shows that the leading term of ${\rm Im} \, \Pi^{\mysmall(2)}(q^2)$ in the limit $\beta \to 0$ is $\alpha \beta/2$, in agreement with~\eq{pithr2}. The function $A(\beta)$, obtained via the nonrelativistic Coulomb Green's function, resums the nonrelativistic vacuum polarization diagrams with the exchange of two or more photons between the electron-positron pair, therefore corresponding to the sum of the leading contributions for $\beta \to 0$ of all vacuum polarization diagrams with three or more loops~\cite{Baier:1969zz}. For $|\beta| \lesssim \alpha$, $A(\beta)$ is of ${\cal O}(\alpha^2)$, whereas for $|\beta| \gg \alpha$ it contains terms of ${\cal O}(\alpha^3)$ and higher, as it can be immediately seen expanding it for $|\beta|>\alpha/2$,
\be
	A(\beta) = \frac{\alpha^2}{2} \sum_{k=1}^{\infty}  \zeta(k+1) \! \left( \frac{i \alpha}{2 \beta} \right)^{\! k} \!.
\label{eq:pithrAexp}
\ee

Remarkably, the function $A(\beta)$ catches the threshold effects both above and below $q^2=4m^2$. In fact, as the digamma function $\psi(z)$ has simple poles at $z=0,-1,-2,\ldots$, $A(\beta)$ has poles at $\beta = i \alpha/2n$ which, to leading order in $\alpha$, correspond to $q^2=M_n^2$, the energy states of positronium. Developing the Laurent expansion of~\eq{pithrA} about the positronium poles and selecting the imaginary part of $\pithr(q^2)$ for all values of $q^2$, one obtains~\cite{Hoang:1997vs}
\bea
	{\rm Im} \,  \pithr(q^2)  &=&  16 \pi^2 \alpha \sum_n \frac{|\phi_{n,0}(0)|^2}{M_n}  \, \delta \! \left(q^2 -M_n^2 \right) 
	\nonumber \\
	  && + \,\,\,  \theta \! \left( \beta \right)   \frac{\pi \alpha^2/2}{1-e^{-\pi \alpha / \beta}}.
\label{eq:impithr}
\eea
The first line of~\eq{impithr} agrees with the contribution of the positronium poles to ${\rm Im} \,\Pi(q^2)$ in~\eq{ImPi}. The second line, which provides the continuum contribution, is the Sommerfeld factor.

With $\pithr(q^2)$ at our disposal, we will now follow an argument similar to one in~\cite{Melnikov:2014lwa} to verify that the total (positronium poles plus continuum) nonperturbative contribution to the electron $g$$-$$2$ arising from the threshold region is equal to $a_e^{\mysmall \rm  P}/2$. Starting from $a_e({\rm vp})$ in~\eq{aevp}, this contribution is given by
\be
	a_e^{\rm\mysmall{thr}}({\rm vp}) = 
	\frac{\alpha}{\pi^2} \int_{M_1^2}^{q_0^2} \, \frac{ds}{4m^2} \,\,
	{\rm Im} \! \left\{A(\beta) -\frac{i \pi^2\alpha^3}{24\beta} \right\} K(4m^2),
\label{eq:aevpthr}
\ee
where $M_1= 2m-{\cal E}_1$ is the energy of the positronium ground state and $q_0^2>4m^2$ corresponds to $\beta_0=\beta(q_0^2)$ with $\pi \alpha \ll \beta_0 \ll 1$.  With these integration limits, $a_e^{\rm\mysmall{thr}}({\rm vp})$ catches the contribution of the entire threshold region. The expression in braces in~\eq{aevpthr} is $\pithr(q^2)$ subtracted of the ${\cal O}(\alpha)$, ${\cal O}(\alpha^2)$ and ${\cal O}(\alpha^3)$ terms of its perturbative expansion (see Eqs.~(\ref{eq:pithr}) and (\ref{eq:pithrAexp})); this subtracted quantity selects the nonperturbative contribution of the threshold region, which arises at ${\cal O}(\alpha^4)$. Equation~(\ref{eq:aevpthr}) can be split into its poles and continuum parts, and, using \eq{impithr}, can be written in the form (note that $\beta$ is imaginary at the poles)
\bea
	\!\!\!\!\!\!\!\!\!
	a_e^{\rm\mysmall{thr}}({\rm vp}) &=& 
	\frac{\alpha}{\pi^2} \, K(4m^2) \left\{ \int_{M_1^2}^{4m^2} \frac{ds}{4m^2}
	\,\, {\rm Im} \, A(\beta) \right.
	\nonumber \\
	&& \!\!\!\!\!\!\!\!\!\!\!\!\!\!\!\!\!\!\!\!\!\!\!\!\!\!\!\!\!\!\!\!
	\left.  + \int_{0}^{\beta_0} \! 2\beta \, d\beta \left[ \frac{\pi \alpha^2/2}{1-e^{-\pi \alpha / \beta}}           
	-\frac{\alpha \beta}{2} -\frac{\pi \alpha^2}{4} -\frac{\pi^2\alpha^3}{24\beta} \right] \right\}\!.
\label{eq:aevpthr2}
\eea
The function $A(\beta)$ has branch points at $q^2=0$ and $4m^2$ and, as discussed above, simple poles at $q^2=M_n^2$. Employing a dispersion relation for the real part of $A(\beta)$,~\eq{aevpthr2} can be expressed in terms of ${\rm Re} \, A(\beta)$ at $|q^2| \to \infty$, i.e.\ $\beta \to 1$. To leading order in $\alpha$ we obtain
\be
	a_e^{\rm\mysmall{thr}}({\rm vp}) = - \frac{\alpha}{\pi}  \, K(4m^2) \,  {\rm Re} \, A(1).
\label{eq:aevpthr3}
\ee
This very simple formula can be immediately evaluated using~\eq{pithrAexp} at leading order. The result is
\be
	a_e^{\rm\mysmall{thr}}({\rm vp}) = \frac{\alpha^5}{8\pi} \zeta(3) \, K(4m^2) = \frac{a_e^{\mysmall \rm  P}}{2}.
\label{eq:aevpthr4}
\ee
This consistency check agrees with Eqs.~(21) and (25) of Ref.~\cite{Melnikov:2014lwa}, and confirms that the total contribution of the threshold region to $a_e({\rm vp})$ is equal to the sum of the poles' contribution in~\eq{aeP} and the continuum one in~\eq{aeC}.

We will now show that the above derived threshold contribution $a_e^{\rm\mysmall{thr}}({\rm vp}) = a_e^{\mysmall \rm  P}/2$ is already included in the usual perturbative QED calculations of Refs.~\cite{Aoyama:2012wj,Baikov:2013ula}. To this end, we use the explicit expressions for $\Pi^{\mysmall(8)} (q^2)$, the QED vacuum polarization function at four loops recently computed in Ref.~\cite{Baikov:2013ula}. The authors provide  expansions for  the low-energy, high-energy and threshold regions. In particular, in the threshold region $\Pi^{\mysmall(8)} (q^2)$ can be written as
\be
	\Pi^{\mysmall (8)} \! (q^2) = \sum_{k=-2}^\infty \Pi^{\mysmall (8)}_k \! (q^2) \, \beta^k.
\label{eq:pi8}
\ee
The five-loop QED contribution to $a_e$ arising from the insertion of the eight-order (four-loop) vacuum polarization in the photon line of the second-order vertex diagram has been computed via the formula~\cite{Lautrup:1971jf,KVP}
\be
	a_e^{\mysmall (10)}({\rm vp}) = 
	- \frac{\alpha}{\pi} \int_0^1 dx \, (1-x) \, \Pi^{\mysmall (8)}  \!\!  \left( - \frac{m^2 x^2}{1-x} \right).
\label{eq:aevpeuclidean}
\ee
If we select the first term in the expansion of $\Pi^{\mysmall (8)} (q^2)$ in powers of $\beta$ given by~\eq{pi8}, and replace in it 
$1/\beta^2 = x^2/[x^2+4(1-x)]$, we obtain
\be
	a_e^{\mysmall (10)}({\rm vp}) = - \frac{\alpha}{\pi} \int_0^1 \! (1-x) \!
	\left[ \frac{x^2 \,  \Pi^{\mysmall (8)}_{-2}}{x^2 +4(1-x)} \right] dx \, + \, \cdots
\label{eq:aevpeuclidean2}
\ee
(we note that the expansion in~\eq{pi8} is not well defined in the integration region of~\eq{aevpeuclidean2}, where $\beta \geq 1$, and it is only employed to isolate the term of ${\cal O}(1/\beta^2)$).
The coefficient $\Pi^{\mysmall (8)}_{-2}$ is constant and given by the explicit calculation of Ref.~\cite{Baikov:2013ula},
\be
	\Pi^{\mysmall (8)}_{-2} \, = \, - \, n_e \frac{\alpha^4}{8} \zeta(3),
\label{eq:Pi8minus2}
\ee
where the label $n_e$ (to be set to one) indicates that this term arises from four-loop diagrams with only one closed electron loop. Inserting~(\ref{eq:Pi8minus2}) into~(\ref{eq:aevpeuclidean2}) we obtain
\be
	a_e^{\mysmall (10)}({\rm vp}) \, = \, n_e \, \frac{\alpha^5}{8\pi} \zeta(3) \, K(4m^2) + \cdots 
	\, = \, \frac{a_e^{\mysmall \rm  P}}{2} + \cdots,
\label{eq:aevpeuclidean3}
\ee
which shows that the contribution $a_e^{\mysmall \rm  P}/2$ is naturally included in the perturbative five-loop calculation. Equation~(\ref{eq:aevpeuclidean3}) also shows that this contribution arises from the five-loop set I(i) of Ref.~\cite{KVP} which contains eighth-order vacuum polarization diagrams with only one closed electron loop. This is at variance with the claim of Ref.~\cite{Hayakawa:2014tla} that the leading-order contribution of positronium to $a_e({\rm vp})$ occurs through diagrams of ${\cal O}(\alpha^7)$ obtained from the five-loop set I(j) by adding the exchange of at least one additional photon in each of the two light-by-light scattering loops.

Finally, from~\eq{pi8} we note that $\pithr^{\mysmall(8)}(q^2)$, the leading term of the four-loop function $\Pi^{\mysmall(8)}(q^2)$ in the limit $\beta \to 0$, is equal to $\Pi^{\mysmall (8)}_{-2}/\beta^2$. From~\eq{Pi8minus2} we see that this explicit result is in agreement with the  ${\cal O}(\alpha^4)$ term of the expansion of $A(\beta)$ in~\eq{pithrAexp}. To leading order in $\alpha$ we can therefore express~\eq{aevpthr3} in the form
\be
	a_e^{\rm\mysmall{thr}}({\rm vp}) = - \frac{\alpha}{\pi}  \, K(4m^2) \,  \pithr^{\mysmall(8)}\left(|q^2| \! \to \!  \infty\right).
\label{eq:aevpthr4}
\ee
This result shows that the contribution of the threshold region can be mapped into one at $|q^2| \to \infty$ where, far from the positronium bound states, perturbation theory converges well. This observation, presented in Ref.~\cite{Melnikov:2014lwa} (where it was introduced via the nonrelativistic Coulomb Green's function in the space-like limit $q^2 \! \to \! -\infty$) led the authors to argue that the term $a_e^{\mysmall \rm  P}/2$ can be obtained through conventional perturbation theory, where loop diagrams are calculated performing a Wick rotation with subsequent integration over space-like momenta. Our~\eq{aevpeuclidean3} shows this point explicitly.

\vspace{7mm}
\section{Conclusions}
\label{sec:Conclusions}

In this paper we re-examined the contribution $a_e^{\mysmall \rm  P}$ of positronium to the electron $g$$-$$2$ computed in Ref.~\cite{Mishima:2013ama}. We confirmed the result of this reference and corrected a few errors in its derivation.

As shown recently in Ref.~\cite{Melnikov:2014lwa}, the integral representation for $a_e({\rm vp})$ also receives a continuum nonperturbative contribution from the integration region right above the electron-positron threshold. This additional nonperturbative contribution was shown in~\cite{Melnikov:2014lwa} to cancel one-half of that of positronium. In order to verify this partial cancellation, we introduced the closed-form QED vacuum polarization function near threshold of Refs.~\cite{Braun1968,BCR1973} and calculated the contribution to $a_e({\rm vp})$ arising from its integration in the region below and above threshold. Our result confirms that the total contribution to $a_e({\rm vp})$ of the region near threshold is equal to $a_e^{\mysmall \rm  P}/2$.

We therefore addressed the question whether this remaining term $a_e^{\mysmall \rm  P}/2$ should be added to the perturbative five-loop QED result of Ref.~\cite{Aoyama:2012wj}. The authors of Ref.~\cite{Melnikov:2014lwa} argued that this term is already included in the perturbative ${\cal O}(\alpha^5)$ contribution to $a_e({\rm vp})$ computed in Ref.~\cite{Aoyama:2012wj} and, therefore, should not be added to it. On the other hand, one of the authors of Ref.~\cite{Aoyama:2012wj} recently claimed that positronium contributes to $a_e({\rm vp})$ only through a class of diagrams of ${\cal O}(\alpha^7)$~\cite{Hayakawa:2014tla}. Using the analytic four-loop vacuum polarization function of Ref.~\cite{Baikov:2013ula}, we showed explicitly that the perturbative five-loop calculation of $a_e({\rm vp})$ of Ref.~\cite{Aoyama:2012wj} indeed includes the remaining term $a_e^{\mysmall \rm  P}/2$. We also showed that this contribution arises from the class I(i) of five-loop diagrams containing only one closed electron loop, thus refuting the claim of Ref.~\cite{Hayakawa:2014tla}.

In conclusion, we showed by explicit calculation that there is no additional contribution of QED bound states to $a_e$ beyond perturbation theory.

\begin{acknowledgments}
We would like to thank G.~Dall'Agata, G.~D'Ambrosio, F.~Feruglio, T.~Gehrmann, M.~Hayakawa, H.~Leutwyler, W.~J.~Marciano, P.~Paradisi and E.~Remiddi for very useful discussions. We are also very grateful to G.~P.~Lepage,  G.~Mishima, M.~ Steinhauser and A.~Vainshtein for valuable correspondence. 
The work of M.~F.\ is supported in part by the European Program LHCPhenoNet (PITN-GA-2010-264564).
M.~P.\ also thanks the Department of Physics and Astronomy of the University of Padova for its support. His work was supported in part by the PRIN 2010-11 of the Italian MIUR
and by the European Program INVISIBLES (PITN-GA-2011-289442).
\end{acknowledgments}


\end{document}